\documentstyle[aps,prl,twocolumn,epsfig]{revtex}
\begin{document}
\draft
\title{Modeling Two-Roton Bound State Formation in Fractional Quantum Hall System}
\author{Tarun Kanti Ghosh and G. Baskaran}
\address
{The Institute of Mathematical Sciences, C. I. T. Campus, Taramani, Chennai 600 113, India.}
\date{\today}
\maketitle

\begin{abstract}

Composite Fermion approach using extensive and parallelized numerical analysis has recently 
established a 
two-roton bound state as the lowest energy long wavelength neutral excitation of fractional
quantum Hall effect for 
finite particle ($ N \sim ~ 30 $ ) system. By focusing on the {\em oriented 
dipole } character of magneto 
roton, we model the two roton problem and solve it variationally (analytically) to find 
a two roton bound state with binding energy which is in good agreement with the composite
Fermion numerical results.

\end{abstract}


\narrowtext

The pioneering work of Girvin, Macdonald and Platzman (GMP) \cite{gir} brought out non-trivial 
inner structure of 
neutral excitations of the Fractional Quantum Hall Effect (FQHE) \cite{tsui} systems. This inner 
structure is very transparent for the magneto roton, the minimum energy neutral 
excitations at a finite wave vector $ k_{0} l_{0} = 1.4 $ for the $ \nu = \frac{1}{3}$ quantum Hall 
state. They are 
well approximated by a Laughlin quasi-hole ($\frac{e}{m}$) - quasi-particle ($-\frac{e}{m}$) bound 
state, as shown in Fig. 1a. The composite Fermion
(CF) \cite{jain} approach, that goes beyond Laughlin hierchy of $\frac{1}{m}$ filling, views 
the neutral 
excitations as a {\em composite Fermion interband excitons } of the `pseudo' Landau bands.
 The CF approach has also suggested variational schemes that is amenable to numerical 
studies. Theoretical studies of neutral excitations have become meaningful in the light 
of the Raman scattering experiments \cite{moon}- \cite{dav}.

The aim of the present Letter is to provide 
 an effective microscopic model that uses the essential  structure of 
magneto roton. In our parameter free theory we get the 
two-roton ( with zero total momentum ) 
binding energy that is in good agreement with the extensive finite particle system study 
result in the CF approach. We first show that the magneto roton is an $ oriented $ 
 $ dipole $ analogous to the description of a magnetic exciton 
\cite{bych}, \cite{laughlin}, \cite{kallin} as 
well as Read's \cite{read} dipole description of the neutral composite fermion 
at $ \nu = \frac{1}{2} $. The dipole moment vector ( $ \vec d $ ), the momentum of the magneto 
roton ($\vec p $ ) and the external magnetic field ($\hat{z}$ ) form a triad 
$ (\hat{ z} = \frac{\vec d \times \vec p }{|\vec d | |\vec p |} $). 
The oriented character of the dipole moment leads to a velocity dependent effective 
interaction between two rotons, analogue to the velocity dependent interaction found by one of the 
author's \cite{bas}, in the context of BCS instability of composite Fermi liquid. The resulting two body 
problem is solved variationally to find a bound state. The binding energy of our ( parameter 
free ) theory is in good agreement with the numerical results of Park and Jain \cite{park}.

In the single mode approximation ( SMA ) a neutral excitation is defined by (unnormalized) 
wave function $ \psi_q = P_{LLL} \rho_q \psi_L $ ,  where $ \rho_q = \sum_i
e^{i\vec q . \vec r_i} $.
The minimum of energy $( E_{min} = E_{rot} )$ occurred at $ k = k_0 $ and this excitation was 
called magneto roton,  by analogy with roton of liquid $^4$helium \cite{fey}.

It was observed by GMP \cite{gir} that the zero momentum neutral excitation,  as observed by 
numerical experiment \cite{hald} was in disagreement with their result at E($ k = 0 $). Since the 
numerically observed results was slightly less than $ 2 E_{rot}( k = k_0 )$, they 
speculate that the minimum energy excitation could be a two-roton bound state, as shown in Fig. 1b. 
Within the Landau-Ginzburg theory, Lee and Zang \cite{lee} also proposed that the $ k = 0 $ excitation
consists of two dipoles (two magneto-rotons ), arranges in such a way that it has quadrupole moment
 but the net dipole moment is zero.
Two-roton bound state are suspected to occur in liquid $^4$helium \cite{ruv}.

In a recent paper, Park and Jain \cite{park} have extended their CF exciton theory of magneto 
roton to two-roton bound state problem. Using parallel computing technique, they
have handled upto 30 particle systems. They have shown very convincingly that the zero momentum lowest 
energy excitations is a two-roton bound state.

According to Laughlin \cite{laugh}, the elementary charged excitations, at the filling fraction $ 
\nu = 
\frac{1}{m}$, are quasiparticles (q.p) and quasiholes (q.h) with fractional charge $ \pm \frac{ 
e}{m} $. The effective magnetic length for a particle  with fractional charge $ 
\frac{e}{m}$ is $ l_0 \sqrt m $, where $ l_0 = \sqrt {\frac{\hbar c}{e B}} $ is the magnetic 
length for a particle with charge e.  A roton with wave vector  $ |\vec k|$ is a bound state
 of a q.p and q.h separated by a large distances $ m k l_{0}^2 $. A q.p and q.h have an 
attractive Coulomb interaction $ V (\vec r) = - \frac{ e^2}{m^2 \epsilon r } $. In lowest Landau 
level at filling fraction $ \nu = \frac{1}{m} $ ( m is an odd integer), they obey the following
 guiding center dynamics  \cite{read}, \cite{bas}:
\begin{equation}
\frac{ d \vec r_e }{dt} = \frac{m l_{0}^2 }{\hbar} \nabla_{e} V(\vec r_{e} - \vec r_h) \times 
\hat z
\end{equation}

\begin{equation}
\frac{ d \vec r_h }{dt} = - \frac{m l_{0}^2 }{\hbar} \nabla_{h} V(\vec r_{e} - \vec r_h) \times 
\hat z
\end{equation}
where $ \vec r_e $ and $ \vec r_h $ are the co-ordinates of the q.p and the q.h. These equation
leads to a drift velocity $ \vec v_d = \frac{d}{d t}\vec R $ of the 
center of mass of the pair: 
\begin{equation}
\vec v_d = \frac{m l_{0}^2}{ \hbar} (\nabla_{\vec r} V(\vec r ) \times \hat z )
\end{equation}
where $ \vec r = \vec r_e - \vec r_h $ is the relative distance between the q.p and q.h. and $ \vec R = 
\frac{( \vec r_e + \vec r_h )}{2} $ center of mass co-ordinate. Since the q.p and q.h carry opposite charges,
they both 
drift in a direction perpendicular to their separation vector ${\vec r}$ 
direction. 
\begin{equation}
 \vec r .\vec v_d = \frac{m l_{0}^2}{ \hbar} \vec r . (\nabla_{\vec r} V(\vec r ) \times \hat z )
  = \frac{m l_{0}^2}{ \hbar} \hat z . (\vec r \times \nabla_{\vec r} V(\vec r ))
=0,
\end{equation} 
since $ \vec r \times \nabla_{\vec r} V(\vec r ) $ is zero. Hence $ \vec r .\vec v_d = 0 $. 
 Laughlin's quasi-exciton wave function ( see Eq. (8) of Ref. \cite{laughlin}) can be re-written 
in terms of the center of mass and the relative co-ordinates as
\begin{eqnarray}
| z_0> & = &  \frac{1}{\sqrt{2 m L}} \int \int e^{i \vec R.(\vec k + \frac{\vec r \times \hat z}{2 m 
l_{0}^2})} e^{-\frac{ |\vec r - m l_{0}^2 \hat z \times \vec k |^2}{4 m l_{0}^2}}
\\ \nonumber & & S_{z_e}^{\dag} S_{z_h} | m > d^2 z_e d^2 z_h
\end{eqnarray}
The amplitude of this wave function is maximum when $ \vec r = m l_{0}^2 \vec k \times \hat z $.
So Laughlin's quasi-exciton wave function strongly suggest the oriented nature of our dipole.
 This dipole dynamics is very similar to the dynamics of a vortex anti-vortex in fluid dynamics.
The distance between the constituent particles of a roton ( {\em oriented dipole} 
) is $ \vec r = m l_{0}^2 (\hat z \times \vec k) $. 
The dipole moment of this roton is $ \vec d = \frac{e}{m} m l_{0}^2 (\hat z \times \vec k )$
or $ \vec d = e l_{0}^2 (\hat z \times \vec k ) $. The dipole moment is the same for $ \nu = 
\frac{1}{3} $ and $ \nu = \frac{1}{2} $ for a given $ \vec k $.

At filling fraction $\nu = \frac{1}{m} $, there is a parabolic dispersion around 
the 
minimum energy at finite $ k = k_0 $. The energy spectrum can be written around the minimum 
energy at $ k = k_0 $ as
\begin{equation}
E(k) = E_{rot} + \frac{\hbar ^2}{2m_r} (| \vec k | - k_0 )^2,
\end{equation}
where $ E_{rot}$ is the minimum roton energy at $ k = k_0 $ 
and $ m_r $ is the roton mass.
The minimum roton energy $ E_{rot} $ and the
corresponding $ k_0 $ are different for different filling fractions. So the kinetic energy of a
roton is different for different filling fractions through $ m_{r} $ and $ k_{0} $.
For $ \nu = \frac{1}{3}$, $ E_{rot} = 0.075 \frac{e^2}{\epsilon l_0} $ is the minimum roton energy at 
$ k_0 l_0 = 1.4 $. $ m_r = 2\epsilon \hbar^2 /e^2 l_0 $ is the roton mass. 
 The mass of a roton  is  calculated from 
the curvature of the excitation spectrum  at $ \nu = \frac{1}{3} $ given in Ref. \cite{gir} by using 
the relation $ m_r = \frac{\hbar ^2}{\frac{\partial ^2 E(k)}{\partial k^2}} $ at $ k = k_0 $.

 The kinetic energy for two roton with momenta $ \vec k_1 $ and $ \vec k_2 $ is
\begin{equation} 
 T = \frac{\hbar^2}{2 m_{r}}[(|\vec k_1| - k_0 )^2 + ( |\vec k_2 | - k_0 )^2].
\end{equation}

Since each roton is an $ oriented $ $ dipole $, it is a natural choice to consider the 
interaction between two roton as a dipole-dipole interaction. This momentum dependent dipole-dipole 
interaction was first suggested by one of the author's \cite{bas} for $ \nu = \frac{1}{2}$ composite
 Fermi liquid.
 The classical dipole-dipole 
interaction energy with two dipoles $ \vec d_1 $ and $ \vec d_2 $ is,
\begin{equation}
U = \frac{\vec d_1 .\vec d_2}{ \epsilon r_{12}^3} - 3 \frac{( \vec d_1 .\vec r_{12})( \vec d_2 . 
\vec r_{12})}{ \epsilon r_{12}^5},
\end{equation}
where $ \vec r_1 $ and $ \vec r_2 $ are the position vectors of the two dipoles and $ \vec 
r_{12} = \vec r_1 - \vec r_2 $ is the relative distance between two dipoles. $ \epsilon $ is 
the dielectric constant of the background material.

 $\vec d_1 = e l_{0}^2(\hat{z}\times \vec k_1)$ and $ \vec d_2 = e l_{0}^2 
(\hat{z}\times \vec k_2) $ are the dipole moments of the two roton with wave vector
 $ \vec k_1 $ and $ \vec k_2$  respectively and $ | \vec r_{12} | $ is the relative distance between two rotons.
Using the dipole moments $ \vec d_1 $ and $ \vec d_2 $ for  the two rotons, this interaction 
energy can be rewritten in terms of the total momentum $ 
\vec K = \vec k_1 + \vec k_2 $ and  relative momentum $ \vec k = \vec k_1 - \vec k_2 $ as
\begin{equation}
U = \frac{ e^2 l_{0}^4}{4 \epsilon} [ \frac{( \vec K ^2 - \vec k ^2 )}{r_{12}^3} - 3 
\frac{(\hat z \times (\vec K + \vec k).\vec r_{12})(\hat z \times 
(\vec K - \vec k).\vec r_{12})}{r_{12}^5}]
\end{equation}
This is a semi classical expression for the potential energy of two interacting {\em oriented dipoles }. 
 Since an {\em oriented dipole} is a quantum mechanical particle, we pass on to 
quantum dynamics by symmetrize 
the above classical energy expression and replace the total momentum $ \vec K $ and the 
relative 
momentum $ \vec k $ by an operator $ - i \nabla_{\vec R } $ and $ - i \nabla_{\vec r} $
respectively. 
After symmetrize, interaction energy reduces to
\begin{eqnarray}
U & = & \frac{ e^2 l_{0}^4}{4 \epsilon} [ \frac{ \vec K^2}{r_{12}^3} - \frac{1}{r_{12}^3} \vec k^2 - 
\vec  k^2 (\frac{1}{r_{12}^3})
\\ \nonumber & - & 3 \frac{(\hat z \times \vec K). \vec r_{12} (\hat z \times \vec K). \vec
 r_{12} - (\hat z \times \vec k). \vec r_{12} (\hat z \times \vec k). \vec r_{12}}{r_{12}^5}] 
\end{eqnarray}
In operator form, it becomes
\begin{eqnarray}
U & = & \frac{e^2 l_{0}^4}{4 \epsilon} [\frac{- \nabla_{\vec R}^2}{ r_{12}^3} 
+ \frac{1}{2} (\frac{1}{r_{12}^3} \nabla_{\vec r }^2 + \nabla_{\vec r }^2(\frac{1}{r_{12}^3}))
 - \frac{3}{r_{12}^5} \frac{\partial ^2 }{\partial \phi ^2}
\\ \nonumber & + & \frac{3}{r^3} ( \sin ^2 (\theta + \phi ) \frac{ \partial ^2}{ \partial R^2} + 
\cos ^2 (\theta + \phi) \frac{1}{R^2} \frac{ \partial ^2}{\partial \theta ^2} \\ \nonumber & + & 
\cos ^2 (\phi - \theta ) 
\frac{1}{R} \frac{ \partial }{\partial R} +  \sin 2(\phi - \theta) \frac{1}{R^2} \frac{ \partial 
}{\partial \theta })], 
\end{eqnarray} 
where $ \phi $ is the angle between $ \vec r $ and $ x $ axis and $ \theta $ is the angle between  
 $ \vec R $ and $ X $ axis.
The term $ \frac{1}{2} (\frac{1}{ r_{12}^3} \nabla_{\vec r }^2 + \nabla_{\vec r}^2(\frac{1}{r_{12}^3})) 
$ in the above expression is due to symmetrization of $ \frac{\nabla_{\vec r}^2}{r_{12}^3}$ term. 
Without symmetrization of this term $ \frac{\nabla_{\vec r}^2}{r_{12}^3}$,
 the interaction does not give the correct binding energy. So quantum 
mechanics play a  crucial role in the interaction between two rotons.
This is a momentum dependent, non-central potential between two {\em oriented 
dipoles} of non-zero total
momentum. This momentum dependent interaction energy is same for all $ \nu = \frac{p}{(2 m p + 1)} $ filling 
fractions, where $m$ and $ p $ are integers.

Since we are interested in the pair formation,  we concentrate only two roton with 
opposite momenta ($ \vec k_1 = - \vec k_2 $), as done in BCS theory. Hence the total 
momentum is zero.
The interaction energy can be written as
\begin{equation}
U = \frac{e^2 l_{0}^4}{4 \epsilon }[\frac{1}{2}(\frac{1}{r_{12}^3} \nabla_{\vec r}^2 + \nabla_{\vec 
r}^2 (\frac{1}{r_{12}^3})) - \frac{3}{r_{12}^5} \frac{\partial^2}{\partial \phi^2 }].
\end{equation}
 
 The Hamiltonian for this two body problem  with the total momentum $ \hbar \vec K = 0 $ becomes
 \begin{eqnarray}
 H & = & \frac{ \hbar ^2 }{4 m_r}(|i \vec \nabla_{\vec r} | -2 k_0)^2 \\ \nonumber & + &
\frac{e^2 l_{0}^4}{4 \epsilon }[\frac{1}{2}(\frac{1}{r_{12}^3} \nabla_{\vec r }^2 + \nabla_{\vec r}^2
(\frac{1}{r_{12}^3})) - \frac{3}{r_{12}^5} \frac{\partial^2}{\partial \phi^2 }].
\end{eqnarray}

We propose a variational wave function 
\begin{equation}
\psi (r) = N r^2 e^{-\alpha r} J_{0}(2k_0 r),
\end{equation}
where N is the normalization constant which is determined by the condition $ \int d^2 r |\psi 
(r)|^2 = 1 $, $ J_0( 2k_0 r ) $ is the zeroth order Bessel function and $  r = |\vec r_{12}|$. 
 $\alpha $ is the variational parameter which can be determined by minimizing the energy 
expectation value.

In  superconductivity, Cooper pair forms at the Fermi surface between two 
electrons with opposite momenta. Similarly, roton pair forms at and near $ k = k_0 $.
The annular region in k-space that contributes to $ \vec K = 0 $ magneto roton bound state is 
shown in Fig.2.
Like Cooper pair wave function, we construct a wave function for the two-roton bound state with
momenta ($ \vec k_0, - \vec k_0 $) which gives $ J_0(2k_{0} r) $ for s-state. $ 2k_0 $ is the 
relative momentum of these two roton.

To calculate the expectation value of the first term 
\newline
 ( kinetic term ) on the right hand 
side of 
this Hamiltonian, we go to the momentum space. The variational wave function in momentum space 
is 
\begin{equation}
\psi (k) =  N^{\prime} \frac{6}{ \alpha^4 } \int_{0}^{2\pi} d \theta F[2 , 2.5; 1; 
- \frac{( k^2 + 4 k_{0}^2 - 4 k k_0 \cos \theta)}{ \alpha ^2 }]
\end{equation}
where $ N^{\prime} $ is the normalization constant.
In momentum space the kinetic energy operator is
$ T = \frac{\hbar ^2}{4 m} ( k - 2 k_0 )^2 $.
The expectation value of the kinetic energy in k-space is 
$ E_1 ( \tilde{ \alpha } ) = 0.125 E_c 
\int d^2 \tilde {k} |\psi (\tilde {k})|^2 ( \tilde {k} - 2.8 )^2 $, 
where $ E_c = \frac{e^2}{\epsilon l_0}$ is the unit of Coulomb energy and $ 
\tilde{\alpha} = \alpha l_0 $ and $ \tilde{k} = k l_0 $ are the dimensionless variables. 

The expectation value of the interaction term ( second term of the Hamiltonian ) is  
$ E_{2}(\tilde{\alpha}) = \frac{0.125 B}{ A } E_c $ where A and B are the following integrals:
\begin{equation}
A = \int_{0}^{\infty} d \tilde{r} \tilde{r}^5 ( J_{0}(2.8 \tilde{r} ))^2 e^{-2 \tilde{\alpha} 
\tilde{r} },
\end{equation}

\begin{equation}
B   =  \int_{0}^{\infty} d \tilde{r} [ 4 - 2 \tilde{r}^2 \tilde{\alpha}^2 + 2 \tilde{\alpha} 
\tilde{r} - 15.68 \tilde{r}^2 ] J_{0}^2(2.8 \tilde{r})
 e^{- 2 \tilde {\alpha} \tilde {r}}.
\end{equation}
 
We are numerically minimizing the energy functional $ E(\tilde{\alpha} ) = E_1 (\tilde{\alpha}) + 
E_2 (\tilde{\alpha}) $ with respect to the variational parameter $\tilde{ \alpha } $.
  The minimum energy for two-roton state is 0.138 $ E_c $ at $\tilde{ \alpha } = 0.41 $ 
whereas $ 2 E_{rot} = 0.15 E_c $ so that the binding energy is $ 0.012 E_c $. 
Park and Jain have found a minimum energy of  0.135 $ E_c $ and hence the binding energy 
is 0.015 $ E_c $. Our binding energy is thus in good agreement with the extensive 
numerical results of Park and Jain \cite{park}. So two rotons with opposite momenta forms a 
bound state.

The root mean square distance between these two
roton  is  $  \sqrt {<r^2>} \sim 6.7 l_0 $
where as the size of a single roton is approximately 4.2 $ l_0 $.

When the total momentum $ \vec K $ of a two-roton bound state increase, the energy 
is also increased.
At $ K \ge K_c $, the two-roton bound state breaks into two rotons.
To get a qualitative idea  that how the excitation spectrum of a bound state goes 
with the total momentum, we use semiclassical approximation. We consider $ \vec k_1 
= \vec k_0 + \vec q $ and $ \vec k_2 = - \vec k_0 $, where $ |\vec q | < |\vec k_0 | $.  
One can easily get the semiclassical 
energy  $ E(K) = E_c [ 0.3125 (K l_0 )^2 - 0.01 ( 1 + 0.77 (K l_0 )^2)] $
 The critical momentum $ K_c $ can be determined 
 by the condition, $ E (K = 0) + E (K) = 2 E_{rot} $. Using this condition,
 we have $ K_c l_0  = 0.22 $. 
The two-roton bound state is not the lowest energy excitation when $ K \ge K_c $.
Expected excitation spectrum of a two-roton bound state and the two roton continuum 
state is shown in 
Fig. 3. and compared with a single roton excitation spectrum which is given in 
Ref. \cite{gir}.  
The effective mass of a two roton bound state is $ M = 0.054 m_e $ which is 75 
percent less than the sum of the two rotons mass.

In conclusion, we identified the magneto roton as an {\em oriented  dipole}. 
 We derived the momentum dependent, non-central interaction energy form between 
two rotons from a classical dipole-dipole interaction energy. Finally we
proposed a wave function for two-roton state and  shown analytically that at 
$ \nu = \frac{1}{3} $ lowest energy excitation is a two-roton bound state 
(with zero total momentum) instead of a single roton.

One of us (G.B) wish to thank Steve Girvin for a discussion.
This research was supported in part by the National Science
Foundation under Grant No. PHY94-07194.

\begin{figure}
\begin{center}
\epsfig{file=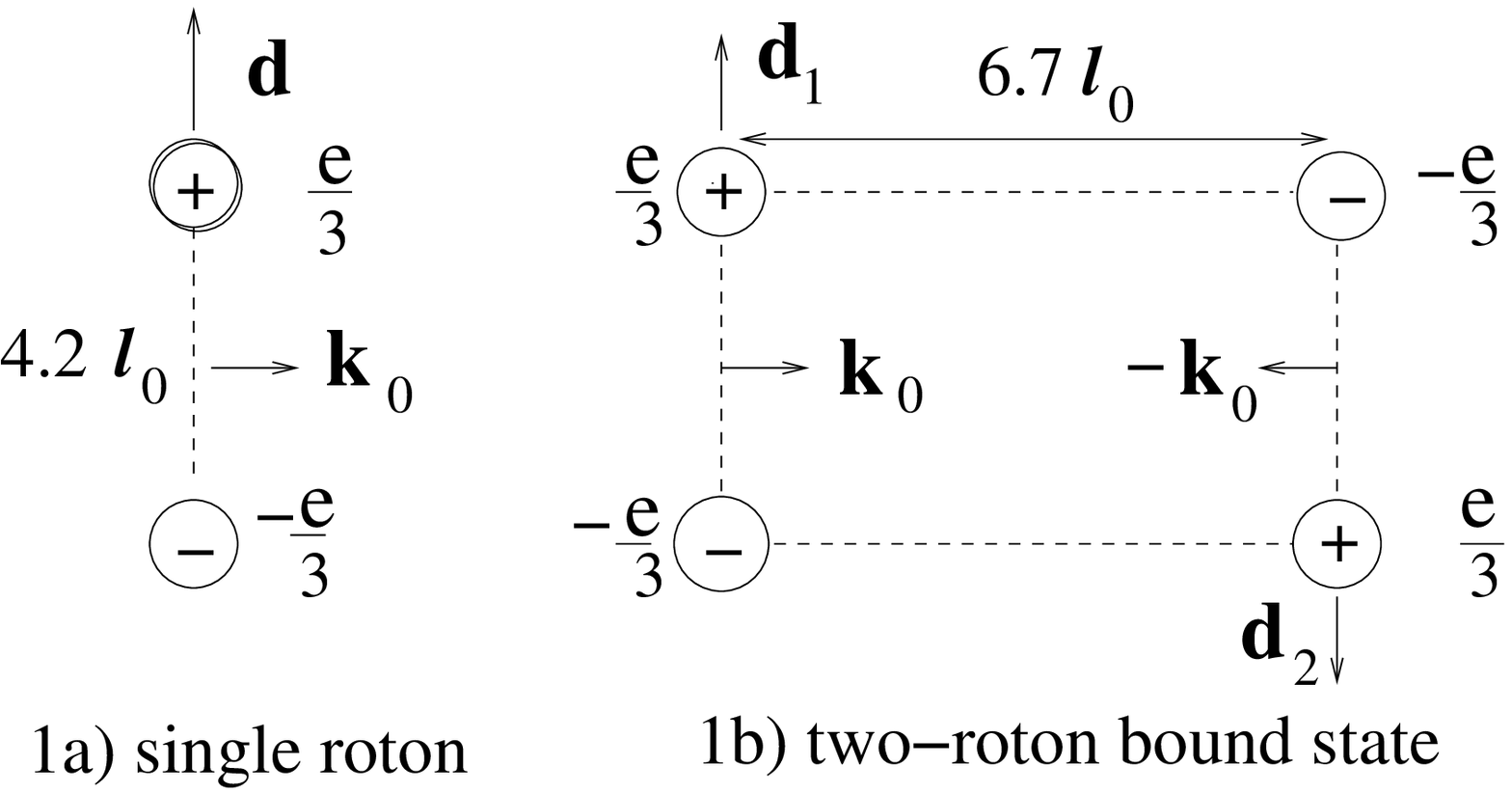, width= 9 cm,angle=0}
\vspace{.5 in}
\begin{caption}
{  Schematic diagrams for ( a ) a single roton with momentum $ \vec k_o  $  and   ( b ) a two-roton bound
state  with  total momentum $ \vec K = 0 $. }
\end{caption}
\end{center}
\label{fig1}
\end{figure}

\begin{figure}
\begin{center}
\epsfig{file=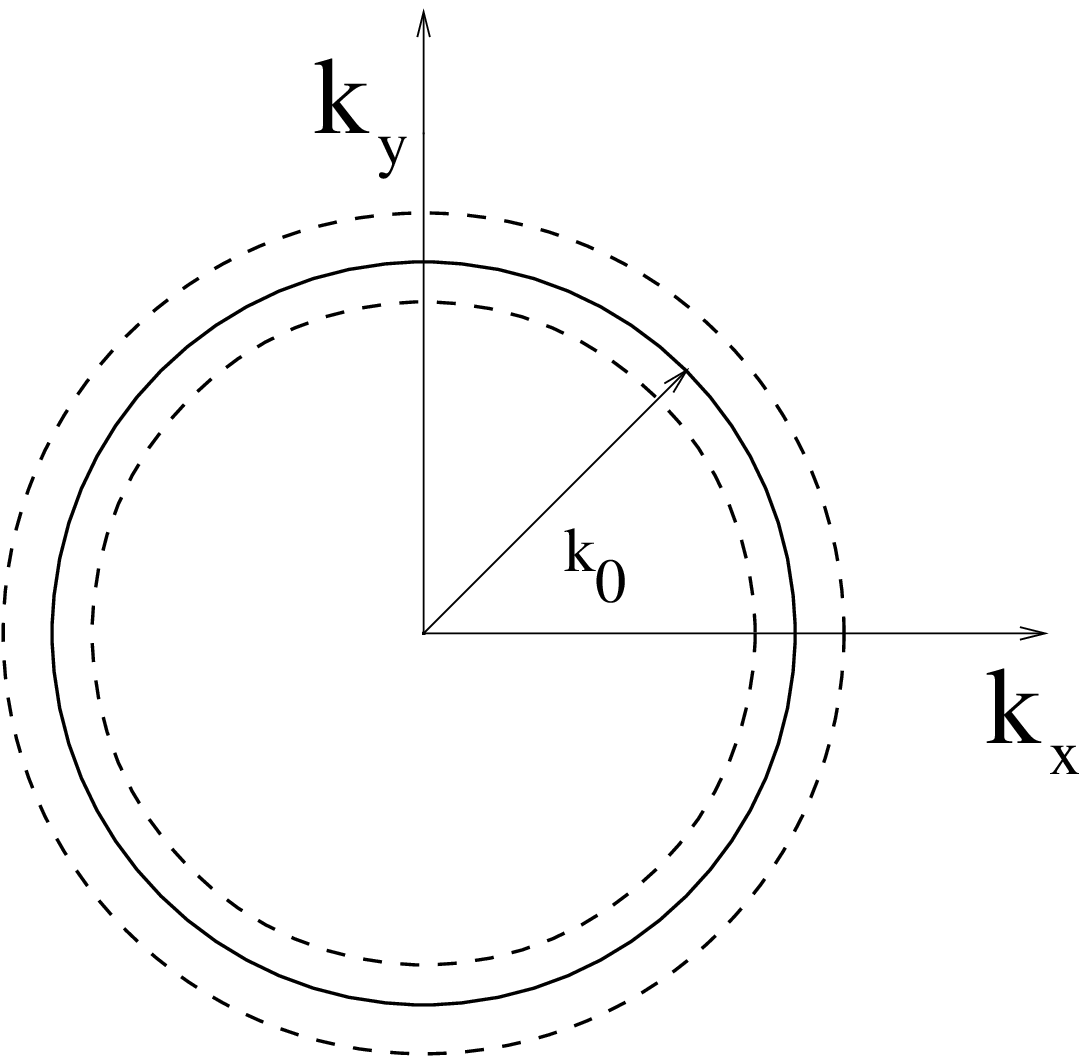, width= 9 cm,angle=0}
\vspace{.5 in}
\begin{caption}
{Annular region in k-space that contributes to $ \vec K = 0 $ magneto-roton   
bound state. }                                                                  
\end{caption}                                                                           
\end{center}  
\label{fig2}
\end{figure}

\begin{figure}
\begin{center}
\epsfig{file=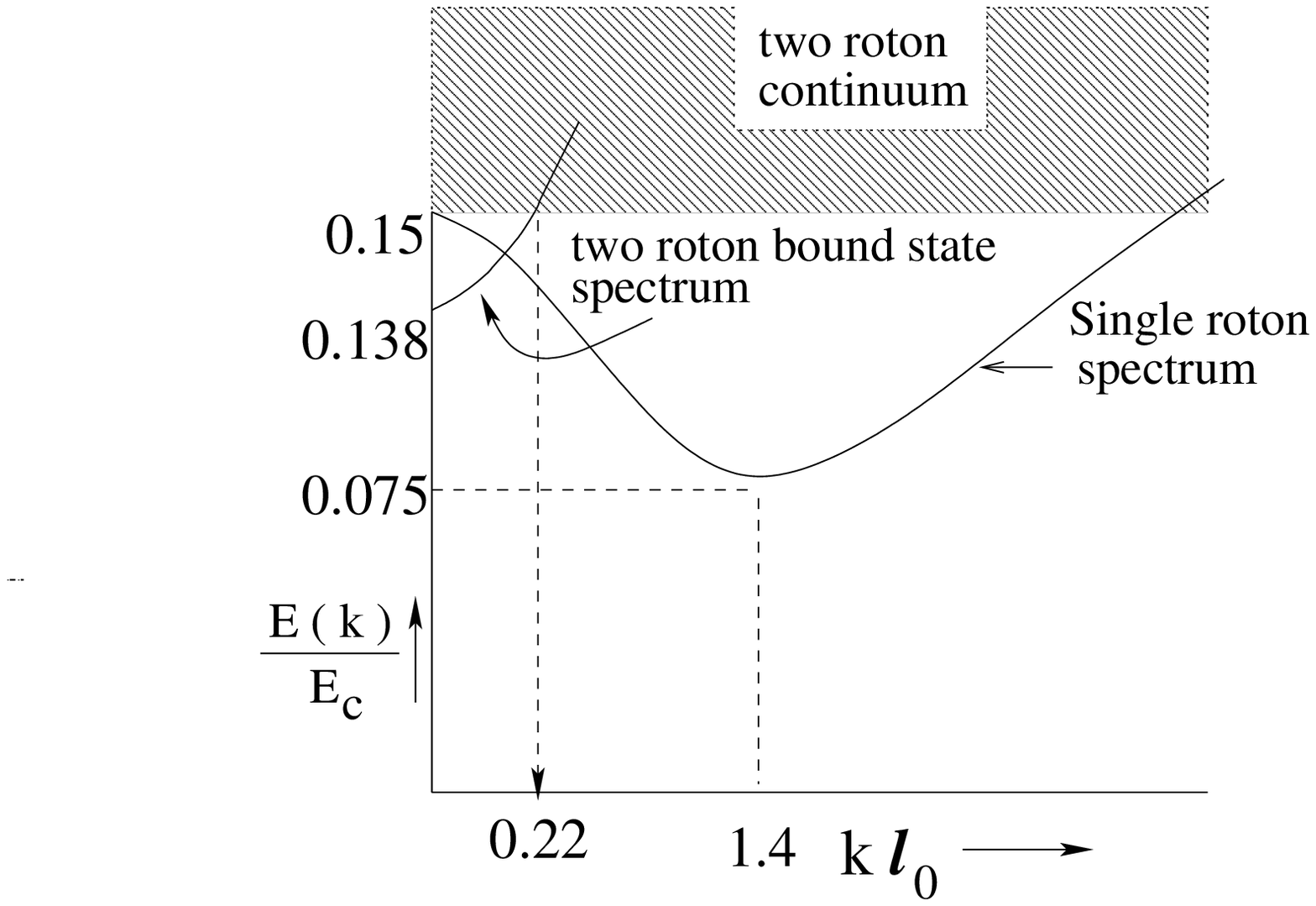, width= 9 cm,angle=0}
\vspace{.5 in}
\begin{caption} 
{Expected qualitative excitation  spectrum  of two-roton bound state  compared with excitation 
spectrum  of a single roton. } \end{caption} 
\end{center}
\label{fig3}
\end{figure}


\begin{thebibliography}{99}
\bibitem{gir}
S. M. Girvin, A. H. Macdonald, and P. M. Platzman, Phys. Rev. Lett. {\bf 54}, 581 (1985); Phys. 
Rev. B {\bf 33 }, 2481 (1986).
\bibitem{tsui}
D. C. Tsui, H. L. Stormer and Gossard, Phys. Rev. Lett. {\bf 48}, 1559 (1982).
\bibitem{jain}
J. K. Jain, Phys. Rev. Lett. {\bf 63}, 199 (1989); Phys. Rev. B {\bf 41}, 7653 (1990).
\bibitem{moon}
M. Kang, A. Pinczuk, B. S. Dennis, M. A. Eriksson, L. N. Pfeiffer, and K. W. West, Phys. Rev. 
Lett. {\bf 84}, 546 (2000).
\bibitem{dav}
H. D. M. Davies, J. C. Harris, J. F. Ryan, and A. J. Turberfield, Phys. Rev. Lett. {\bf 78}, 4095 (1997).
\bibitem{bych} 
Yu. A. Bychkov, S. V. Iordanskii and G. M. Eliashberg, Sov. Phys. JETP Lett. {\bf
33}, 143 (1981) 
\bibitem{laughlin}
R. B. Laughlin, Physica B, {\bf 126}, 254 (1984).
\bibitem{kallin}
C. Kallin and B. I. Halperin, Phys. Rev. B {\bf 30}, 5655 (1984).
\bibitem{read}
N. Read, Semi. Sci. Tech. {\bf 9 }, 1859 (1994); Surf. Sci. {\bf 361/362}, 7 (1996). 
\bibitem{bas}
G. Baskaran, Physica B, {\bf 212}, 320 (1995).
\bibitem{park}
K. Park and J. K. Jain, Phys. Rev. Lett. {\bf 84}, 5576 (2000).
\bibitem{fey}
R. P. Feynman, Statistical Mechanics (Benjamin, reading, Mass, 1972); Phys. Rev. {\bf 91}, 
1291, 1301 (1953); {\bf 94}, 262, (1954); R. P. Feynman and M. Cohen, ibid. {\bf 102}, 1189 (1956).
\bibitem{hald}
F. D. M. Haldane and E. H. Rezayi, Phys. Rev. Lett. {\bf 54}, 237 (1985).
\bibitem{lee}
D. H. Lee and S. C. Zang, Phys. Rev. Lett, {\bf 66 }, 1220 (1991).

\bibitem{laugh}
R. B. Laughlin, Phys. Rev. Lett. {\bf 50}, 1395, (1983).
\bibitem{ruv}
A. Zawadowski, J. Ruvalds, and J. Solana, Phys. Rev. A {\bf 5}, 399 (1972); V. Celli, and J. 
Ruvalds, Phys. Rev. Lett. {\bf 28}, 539 (1972).

\end{thebibliography}
\end{document}